\documentclass[11pt]{article}%
\usepackage{amsfonts}
\usepackage{amsmath}
\usepackage{amssymb}
\usepackage{graphicx}
\usepackage{fullpage}
\usepackage{hyperref}%
\providecommand{\U}[1]{\protect\rule{.1in}{.1in}}
\numberwithin{equation}{section}
\begin{document}

\title{\textbf{Simulations of closed timelike curves}}
\author{Todd A. Brun\thanks{Ming Hsieh Department of Electrical Engineering, Center
for Quantum Information Science and Technology, University of Southern
California, Los Angeles, California, 90089-2565}
\and Mark M. Wilde\thanks{Hearne Institute for Theoretical Physics, Department of
Physics and Astronomy, Center for Computation and Technology, Louisiana State
University, Baton Rouge, Louisiana 70803, USA}}
\maketitle

\begin{abstract}
Proposed models of closed timelike curves (CTCs) have been shown to enable powerful information-processing protocols.  We examine the simulation of models of CTCs both by other models of CTCs and by physical systems without access to CTCs. We prove that the recently proposed transition probability CTCs (T-CTCs)  are physically equivalent to postselection CTCs (P-CTCs), in the sense that one model can simulate the other with reasonable overhead. As a consequence, their information-processing capabilities are equivalent. We also describe a method for quantum computers to simulate Deutschian CTCs (but with a reasonable overhead only in some cases).  In cases for which the overhead is reasonable, it might be possible to perform the simulation in a table-top experiment. This approach has the benefit of resolving some ambiguities associated with the equivalent circuit model of Ralph \textit{et al}. Furthermore, we provide an explicit form for the state of the CTC system such that it is a maximum-entropy state, as prescribed by Deutsch.

\end{abstract}

\section{Introduction}

Simulating one physical system using another is one of the core methods that we employ to understand physics. For example, the use of classical computers to simulate physical systems is ubiquitous nowadays, and one of the great hopes for quantum computation is that quantum computers could reduce the overhead needed to simulate quantum physical systems. Thus, it is of interest to determine how to simulate models of closed timelike curves (CTCs) in order to improve our understanding of these models. 

CTCs are a tantalizing possibility not ruled out by the laws of physics, and they have been the subject of debate for some time now, especially since since the works in \cite{RevModPhys.21.447,B80,PhysRevLett.66.1126}. Beginning with the work of Deutsch \cite{PhysRevD.44.3197}, the quantum information community has entered the discussion.  The common approach has been to strip away details of the underlying spacetime geometry in order to focus on resolving the paradoxes associated with time travel, such as the well known {\it grandfather} and {\it unproven theorem} paradoxes. There are now several models under active consideration, including the original proposal of Deutsch \cite{PhysRevD.44.3197}, post-selected quantum teleportation (attributed to a variety of sources \cite{B05,S09,S11,LMGGS10,Lloyd2010}, see also the discussion in \cite{J04}), and most recently transition probability CTCs \cite{A14}. In what follows, we refer to these models as D-CTCs, P-CTCs, and T-CTCs, respectively.

Treatments of CTCs in the quantum information literature generally distinguish two subsystems:  the {\it chronology-violating system}, which passes through the CTC and returns to its own past, and the {\it chronology-respecting system}, which does not itself pass through the CTC but can interact with the chronology-violating subsystem.  The CTC itself is assumed to have finite duration:  it comes into existence at a particular time (in the local rest frame of the experiment) and ends at a later time. The chronology-respecting system is prepared in some initial state before the CTC begins, and any measurements are deferred until after the end of the CTC.  (These details are not essential but simplify the analysis.)

D-CTCs are based on the following idea: for a given initial state of the chronology-respecting system and a given unitary evolution to take place between the chronology-respecting and -violating systems, nature sets the state of the chronology-violating system to be a fixed point of the evolution to avoid the aforementioned causality paradoxes. The resulting evolution is nonlinear:  the state of the chronology-violating system depends on the initial state of the chronology-respecting system as well as the unitary interaction.

P-CTCs resolve causality paradoxes in a different way, by modeling a CTC as a noiseless quantum channel into the past via quantum teleportation.  Normal quantum teleportation involves a measurement with a random outcome \cite{PhysRevLett.70.1895}, which must be communicated from the sender to the receiver, and hence can only arrive after it has been sent.  In P-CTCs, one outcome occurs with certainty, and the state can be received before it is sent.  Nonlinear evolution arises here as well because the model involves a projection and a renormalization.

Finally, T-CTCs represent a new model of CTCs proposed recently \cite{A14}, with the aim of having fewer problematic features than D-CTCs or P-CTCs while still being somewhat plausible.
The outcome is that they resolved a technical issue with P-CTCs, in which it is unclear what happens if there is no overlap between the state at the output of the unitary evolution and the projection. In the T-CTC model, the state of the CTC is initialized to a random pure state with a weight that is proportional to the probability that the evolution is self-consistent. This approach resolves the ``null projection'' issue inherent with P-CTCs.

Several researchers have argued that D-CTCs would offer great information-processing power. In particular, a party equipped with D-CTCs would have computational power equivalent to the computational complexity class PSPACE (which contains NP) \cite{AW09}\ and would be able to violate the uncertainty principle \cite{BHW09}, the Holevo bound \cite{BHW09}, and the cherished no-cloning theorem \cite{BWW13}. In fact, recent work in \cite{YATHVRLWG14} has shown that the latter violations would be possible using so-called open timelike curves, in which it is not even necessary for an interaction to take place between the systems entering and emerging from wormholes \cite{PhysRevLett.110.060501}.\ Bennett \textit{et al}.~have contested these results \cite{BLSS09}, taking a particular adversarial model of computation and information processing to argue that D-CTCs would offer no added benefit for these tasks. This debate is ongoing \cite{CM10,CMP12}.

P-CTCs are seemingly not as powerful as D-CTCs, but still offer information-processing power far beyond standard quantum mechanics. In particular, Lloyd \textit{et al}.~have proven that a P-CTC-equipped party would have computational power equivalent to the computational complexity class PP \cite{Lloyd2010}, a complexity class conjectured to be less powerful than PSPACE, but also containing NP. (The proof makes use of the results in \cite{A05}.) P-CTCs would also enable perfect distinguishability of an arbitrary set of linearly independent quantum states \cite{BW12}, allowing violation of the uncertainty principle. P-CTCs are weaker than D-CTCs, in that they cannot violate the no-cloning theorem \cite{BW12} and are limited to distinguishing a linearly independent set of states rather than an arbitrary set of states \cite{BHW09}.

An intriguing aspect of the work of Lloyd \textit{et al}~\cite{LMGGS10,Lloyd2010} is that they conducted an experiment which provided a physical simulation of P-CTCs. Of course, they did not actually send photons back in time, but instead simulated postselected teleportation, in which they performed ordinary quantum teleportation, but discarded 3/4 of the experimental data so that the remaining data was consistent with the evolution that would occur if postselected teleportation were possible.
The physical simulation of generic channels from the future to the past was then studied in \cite{PhysRevA.85.022330}, which related the maximum probability of successful simulation with the amount of information transmitted by the channel.
After these works, a different group performed an experiment to simulate D-CTCs \cite{RBMWR14}. The method to do so is admittedly simple:  just compute a fixed point for the given initial state and unitary interaction, prepare the ``CTC system'' in this state, apply the unitary, and discard the ``CTC system.'' Of course, this approach is only feasible for small systems, because computing a fixed point of a given quantum circuit is a PSPACE-complete problem \cite{AW09} (which is why D-CTCs have such vast computational power).

This idea of CTC simulation points to two questions:  \textit{How can we simulate CTCs using table-top experiments?}  And: \textit{Can a given model of CTCs physically simulate a different model of CTCs?} In this paper, we address these questions in two specific ways:
\begin{enumerate}
\item We prove that T-CTCs are physically equivalent to P-CTCs, in the sense that one model can simulate the other with arbitrarily good accuracy and reasonable overhead. This resolves an open question from the original paper on T-CTCs \cite{A14} and establishes that the information-processing capabilities of T-CTCs are exactly those of P-CTCs. As a result, the computational power of T-CTCs is equivalent to that of the complexity class PP.

\item We also show how quantum computers can simulate D-CTCs if many copies of the chronology-respecting input state are available. Our method consists of initializing a large number of control qubits in a superposed state and a large number of system registers in the chronology-respecting input state (hence the need for many copies). We then iteratively apply a controlled version of the unitary interaction such that the CTC system eventually relaxes to a fixed point of the evolution. A final application of the unitary interaction followed by a partial trace over the CTC system leads to a simulation of a D-CTC. The purpose of this latter contribution is two-fold: 1) to offer an arguably more interesting method of simulating D-CTCs other than solving directly for the fixed point of the circuit, and 2) to clarify some concerns about the equivalent circuit model of D-CTCs \cite{RM10,RD12} pointed out in \cite{A14}. We also provide an explicit form for the initial state of the CTC system such that it is a maximum entropy fixed-point state according to Deutsch's maximum entropy postulate.
\end{enumerate}

The rest of the paper proceeds as follows. The next section reviews the three models of CTCs:  P-CTCs, T-CTCs, and D-CTCs. We then present the results described above in detail.  We conclude with a brief summary and some open questions for future work.

\section{Preliminaries}

\subsection{Postselected CTCs}

We briefly review the P-CTC model \cite{B05,S09,S11,LMGGS10,Lloyd2010}. The chronology respecting system $S$ is initialized to a density matrix $\rho_{S}$, and $S$ interacts with the chronology-violating system $C$ by a unitary $U_{SC}$. In the P-CTC model, the system $C$ initially holds one half of a maximally entangled state, with the other half being held by a fictitious system $C^{\prime}$, which is used to induce a noiseless quantum channel into the past. Let $\left\vert \Phi \right\rangle_{CC'}$ denote this maximally entangled state:
\begin{equation}
\left\vert \Phi \right\rangle_{CC'} \equiv \frac{1}{\sqrt{d}} \sum_i 
\left\vert i \right\rangle_{C} \otimes \left\vert i \right\rangle_{C'},
\end{equation}
where $d$ is the Schmidt rank of $\left\vert \Phi \right\rangle_{CC'}$.  $S$ interacts with $C$ by the unitary $U_{SC}$, and finally the state is projected onto $\vert \Phi\rangle \langle \Phi\vert_{CC^{\prime}}$ and renormalized. The output is thus
\begin{equation}
\frac{1}{N}\langle \Phi\vert _{CC^{\prime}}U_{SC}
\left(  \rho_{S}\otimes\vert \Phi\rangle \langle \Phi\vert_{CC^{\prime}}\right)
U_{SC}^{\dag}\vert \Phi\rangle _{CC^{\prime}}
= \frac{1}{N} B_{S}\rho_{S}B_{S}^{\dag},
\label{eq:P-CTC-evolve}
\end{equation}
where $B_{S}\equiv\operatorname{Tr}_{C}\left\{  U_{SC}\right\}$, and one can calculate the normalization factor $N$ as
\begin{equation}
N=\text{Tr}\!\left\{  B_{S}^{\dag}B_{S}\rho_{S}\right\}
.\label{eq:p-ctc-normalization}%
\end{equation}
It is unclear what should happen in the case when the projection is null. One often-stated argument is that this would never happen in practice, due to slight noisy deviations from ideal evolution; states with null projection form a set of measure zero.  
Or perhaps the initial state of the universe is chosen such that null projections never occur.
But this gap does seem to be an imperfection in the P-CTC model.

\subsection{Transition probability CTCs}

We now review the model of CTCs known as {\it transition probability CTCs} (T-CTCs), established in \cite{A14}. The T-CTC model supposes that the chronology-respecting (CR) system $S$\ is prepared in a pure state $\vert \psi\rangle _{S}$ and then interacts with a chronology-violating (CV) system $C$ according to some unitary $U_{SC}$. One assumption of the model is that the CTC system $C$ is prepared in some unknown state $\vert \phi\rangle _{C}$ before interacting with $S$. The state after the interaction is
\begin{equation}
U_{SC}\vert \psi\rangle _{S}\otimes\vert \phi\rangle_{C}.
\end{equation}
The probability that the final state of $C$ is consistent with its initial state is
\begin{equation}
\left\Vert \langle \phi\vert _{C}U_{SC}\vert \psi\rangle_{S}\otimes\vert \phi\rangle _{C}\right\Vert _{2}^{2} ,
\label{eq:consistency-probability}
\end{equation}
where the ``probability of consistency'' is taken to mean the probability that a measurement after the interaction would find $C$ in the same state as it started in.  \cite{A14} then argues that the probability of preparation---i.e., the probability that the CTC begins in the state $\vert \phi\rangle _{C}$---should be proportional to this probability of self-consistency. Thus, in this way, states that are more likely to be self-consistent are more likely to be prepared and vice versa.  In the case that system $C$ is prepared in the state $\vert\phi\rangle _{C}$, the state after the interaction is projected and renormalized to
\begin{equation}
\frac{\langle \phi\vert _{C}U_{SC}\vert \psi\rangle_{S}
\otimes\vert \phi\rangle _{C}}{\left\Vert \langle\phi\vert _{C}U_{SC}
\vert \psi\rangle _{S}\otimes\vert\phi\rangle _{C}\right\Vert _{2}}.
\end{equation}
Finally, since the state of the CTC is unknown, the T-CTC model supposes that it is chosen randomly.  States could be chosen uniformly according to the Haar measure, but some of these states have zero probability of being self-consistent:  i.e., the quantity in (\ref{eq:consistency-probability}) could be equal to zero for some $\vert \phi\rangle _{C}$, so that such choices have zero ``preparation measure.'' So, by the above reasoning (that the probability of preparation is proportional to the probability of self-consistency), the preparation probability measure is biased away from the Haar measure and taken to be
\begin{equation}
d\mu(  \phi)  \equiv\frac{d\phi\ \left\Vert \langle\phi\vert _{C}U_{SC}
\vert \psi\rangle _{S} \otimes\vert\phi\rangle _{C}\right\Vert _{2}^{2}}{\int d\phi\ \left\Vert\langle \phi\vert _{C}U_{SC}\vert \psi\rangle_{S}\otimes\vert \phi\rangle_{C}\right\Vert _{2}^{2}},
\end{equation}
where $d\phi$ represents the Haar measure. Thus, the ensemble resulting at the output of the interaction is given by
\begin{equation}
\left\{  d\mu(  \phi)  ,\frac{\langle \phi\vert_{C}U_{SC}\vert \psi\rangle_{S}\otimes\vert \phi\rangle_{C}}{\left\Vert \langle \phi\vert _{C}U_{SC}\left\vert\psi\right\rangle _{S}\otimes\vert \phi\rangle _{C}\right\Vert_{2}}\right\}  ,
\end{equation}
and the density operator corresponding to this ensemble is
\begin{align}
\left(  \rho_{f}\right)  _{S} &  =\int d\mu(  \phi)
\frac{\langle \phi\vert _{C}U_{SC}\left(  \vert \psi\rangle\langle \psi\vert_{S}
\otimes\vert \phi\rangle\langle \phi\vert _{C}\right) 
U_{SC}^{\dag}\vert\phi\rangle _{C}}{\left\Vert \langle \phi\vert_{C}
U_{SC}\vert \psi\rangle _{S}\otimes\vert \phi\rangle_{C}
\right\Vert _{2}^{2}}
\label{eq:T-CTC-DM}\\
&  =\frac{1}{Z}\int d\phi\ \langle \phi\vert _{C}U_{SC}
\left(\vert \psi\rangle \langle \psi\vert _{S}\otimes
\vert \phi\rangle \langle \phi\vert _{C}\right)
U_{SC}^{\dag}\vert \phi\rangle _{C}\\
&  =\frac{1}{Z}\int d\phi\ \left[  \langle \phi\vert _{C}
U_{SC}\vert \phi\rangle _{C}\right]  \left(  \left\vert
\psi\right\rangle \langle \psi\vert _{S}\right) 
\left[ \langle \phi\vert _{C}U_{SC}^{\dag}\vert \phi\rangle_{C}\right]  ,
\end{align}
with
\begin{equation}
Z\equiv\int d\phi\ \langle \psi\vert _{S}\left[  \left\langle
\phi\right\vert _{C}U_{SC}^{\dag}\vert \phi\rangle _{C}\right]
\left[  \langle \phi\vert _{C}U_{SC}\vert \phi\rangle
_{C}\right]  \vert \psi\rangle _{S}.
\end{equation}

\cite{A14} has proven that the density matrix in (\ref{eq:T-CTC-DM}) is equal to
\begin{equation}
\left(  \rho_{f}\right)  _{S}=\frac{1}{z}\left(  \text{Tr}_{C}\left\{U_{SC}\right\}  
\vert \psi\rangle \langle \psi\vert_{S}
\text{Tr}_{C}\left\{  U_{SC}^{\dag}\right\}  +\text{Tr}_{C}
\left\{U_{SC}\left(  \vert \psi\rangle \langle \psi\vert_{S}
\otimes I_{C}\right)  U_{SC}^{\dag}\right\}  \right)  ,
\end{equation}
where $z$ is a normalization factor.

In the above, if the input is the result of inputting a quantum state $\rho_{S}$ which is an improper mixture, then the output density matrix is given by
\begin{equation}
\left(  \rho_{f}\right)  _{S}=\frac{1}{z}\left(  \text{Tr}_{C}\{U_{SC}\}  
\rho_{S}\text{Tr}_{C}\{  U_{SC}^{\dag}\}
+\text{Tr}_{C}\!\left\{  U_{SC}\left(  \rho_{S}\otimes I_{C}\right)
U_{SC}^{\dag}\right\}  \right)  ,
\label{eq:T-CTC-evolve}
\end{equation}
where, again, $z$ is a normalization factor.

\subsection{Deustchian CTCs}

\label{sec:D-CTC-review}The Deutsch model of CTCs \cite{PhysRevD.44.3197} is rather different from the two models discussed above. The evolution is specified by two objects:\ the initial density operator $\rho_{S}$\ for the chronology-respecting system $S$, and the interaction unitary $U_{SC}$ between system $S$ and the chronology-violating system $C$. The model assumes that the systems $S$ and $C$ are initially in a tensor-product state, presumably because they do not interact until the system $C$ emerges from the past mouth of its wormhole. So the initial state after the CTC system emerges but before any interaction occurs is $\rho_{S}\otimes\sigma_{C}$, where $\sigma_{C}$ is a density operator for the CTC system that we will soon specify. The systems $S$ and $C$ interact according to a unitary $U_{SC}$, leading to the state after the interaction
\begin{equation}
U_{SC}\left(  \rho_{S}\otimes\sigma_{C}\right)  U_{SC}^{\dag}.
\end{equation}
The reduced density matrix of the chronology-respecting system after the interaction is
\begin{equation}
\text{Tr}_{C}\!\left\{  U_{SC}\left(  \rho_{S}\otimes\sigma_{C}\right) U_{SC}^{\dag}\right\}  .
\end{equation}
Let $\mathcal{N}_{U,\rho}$ denote the following quantum channel (a linear completely positive trace-preserving map):
\begin{equation}
\mathcal{N}_{U,\rho}(  \omega_{C})  \equiv\text{Tr}_{S}\!\left\{
U_{SC}\left(  \rho_{S}\otimes\omega_{C}\right)  U_{SC}^{\dag}\right\}  .
\end{equation}
That this map is a quantum channel follows from the fact that it is a concatenation of three quantum channels:  the first appends system $S$ in state $\rho_{S}$ to system $C$; the next is a unitary interaction $U_{SC}$; and the last discards system $S$. To determine the state of the CTC system, Deutsch has postulated that nature initializes it to be a fixed point of $\mathcal{N}_{U,\rho}$ \cite{PhysRevD.44.3197}; that is, a state $\sigma_{C}$ such that
\begin{equation}
\mathcal{N}_{U,\rho}(  \sigma_{C})  =\sigma_{C}.
\end{equation}
This criterion ensures that grandfather paradoxes are ruled out.  It also has the rather strange implication that, in a chronology-violating region, nature knows the state of a quantum system in a different location and also is ``aware of'' the unitary interaction that is about to take place, and then initializes the state of the CTC system based on this information.  (Such effects, while strange, are perhaps not unexpected when information can travel backwards in time.)

Deutsch showed that every quantum channel has at least one fixed point in the form of a density operator \cite{PhysRevD.44.3197}. In fact, the following quantum channel projects onto the fixed points of $\mathcal{N}_{U,\rho}$:
\begin{equation}
\overline{\mathcal{N}_{U,\rho}^{\infty}}(  \omega_{C})
=\lim_{N\rightarrow\infty}\overline{\mathcal{N}_{U,\rho}^{N}}\left(\omega_{C}\right)  ,
\end{equation}
where
\begin{equation}
\overline{\mathcal{N}_{U,\rho}^{N}}(  \omega_{C})  \equiv\frac
{1}{N+1}\sum_{i=0}^{N}\mathcal{N}_{U,\rho}^{i}(  \omega_{C})
,\label{eq:fixed-point-approx}%
\end{equation}
and $\mathcal{N}_{U,\rho}^{i}$ is defined to be $i$ consecutive applications of the channel $\mathcal{N}_{U,\rho}$. From the theory of operator algebras, the fixed-point set of the CPTP map $\mathcal{N}_{U,\rho}$ forms a *-subalgebra \cite{HJPW04} (see also \cite{M05,Wolf12}). This implies that there exists a decomposition of the CTC Hilbert space $\mathcal{H}_{C}$ into a direct sum of tensor products
\begin{equation}
\mathcal{H}_{C}=\bigoplus\limits_{j}\mathcal{H}_{C_{L_{j}}}\otimes
\mathcal{H}_{C_{R_{j}}},
\end{equation}
such that any fixed-point density operator $\sigma_{C}$ of $\mathcal{N}_{U,\rho}$ can be written with respect to this decomposition as
\begin{equation}
\sigma_{C}=\bigoplus\limits_{j}q_{\sigma}(j)\ \sigma_{j}\otimes\rho_{j},
\end{equation}
where $q_{\sigma}$ is a probability distribution that depends on $\mathcal{N}_{U,\rho}$ and the particular fixed point $\sigma_{C}$, $\left\{  \sigma_{j}\right\}  $ is a set of states that depends on $\mathcal{N}_{U,\rho}$ and the particular fixed point $\sigma_{C}$, and $\left\{  \rho_{j}\right\}  $ is a set of states that depends on the map $\mathcal{N}_{U,\rho}$ but is independent of any particular fixed point $\sigma_{C}$. Finally, the action of the fixed-point projection map $\overline{\mathcal{N}_{U,\rho}^{\infty}}$ on an arbitrary state $\omega_{C}$ is given in terms of this decomposition by
\begin{equation}
\overline{\mathcal{N}_{U,\rho}^{\infty}}(  \omega_{C})
=\bigoplus\limits_{j}\text{Tr}_{C_{R_{j}}}\{\Pi_{j}\omega_{C}\Pi_{j}\}
\otimes\rho_{j},
\label{eq:fixed-point-projection}
\end{equation}
where $\Pi_{j}$ is the projection onto the subspace $\mathcal{H}_{C_{L_{j}}}\otimes\mathcal{H}_{C_{R_{j}}}$. The above facts will allow us to address several important points regarding D-CTCs.

The equivalent circuit model from \cite{RM10,RD12} claimed that the map $\lim_{i\rightarrow\infty}\mathcal{N}_{U,\rho}^{i}(  \omega_{C})$ converges to a fixed point of $\mathcal{N}_{U,\rho}$ and that one reaches a maximum entropy fixed point by considering $\lim_{i\rightarrow\infty}\mathcal{N}_{U,\rho}^{i}\left(  \pi_{C}\right)$, where $\pi_{C}$ is the maximally mixed state. However, the existence of a fixed point does not in itself imply that repeated iterations of the map converge to that fixed point.  It was pointed out in \cite{A14} that this need not be the case, and an explicit counterexample was given. We will discuss this question in Section~\ref{sec:D-CTC-simulation}. By invoking the above facts about fixed points, and providing a suitable definition of convergence, we can clarify the situation and show that indeed there is an equivalent circuit that converges to the D-CTC evolution.

\section{P-CTCs and T-CTCs are physically equivalent}

In this section, we prove that P-CTCs and T-CTCs are physically equivalent to each other. That is, each model can simulate the other up to an arbitrary accuracy, and with reasonable overhead.

\subsection{P-CTCs can simulate T-CTCs}

First we show that P-CTCs can simulate T-CTCs, using the evolution equation for T-CTCs from (\ref{eq:T-CTC-evolve}). Our goal is to exploit a P-CTC such that the final state is equal to the state in (\ref{eq:T-CTC-evolve}). To this end, we first prepare three chronology-respecting quantum registers $S$, $C$, and $C^{\prime}$ in the following state:
\begin{equation}
\rho_{S}\otimes\left[  p\Phi_{CC^{\prime}}+\left(  1-p\right)  \pi_{C} \otimes\pi_{C^{\prime}}\right]  ,
\end{equation}
where $\Phi_{CC^{\prime}}$ is the maximally entangled state, $\pi$ is the maximally mixed state, and $p\in(0,1)$. We then apply the unitary $U_{SC}$ to registers $S$ and $C$, leading to the state
\begin{equation}
pU_{SC}\left(  \rho_{S}\otimes\Phi_{CC^{\prime}}\right)  U_{SC}^{\dag}
+\left(1-p\right)  U_{SC}\left(  \rho_{S}\otimes\pi_{C}\otimes\pi_{C^{\prime}}\right)  U_{SC}^{\dag}.
\end{equation}

Now suppose that we have a chronology-violating system consisting of a single-qubit register $D$ that passes through a P-CTC.  We then apply the following controlled unitary:
\begin{equation}
\Phi_{CC^{\prime}}\otimes I_{D}+\left(  I_{CC^{\prime}}
-\Phi_{CC^{\prime}}\right)  \otimes X_{D},
\end{equation}
where $X_{D}$ is a Pauli bit-flip operator. The effect of this unitary together with the P-CTC is to retain the projection onto $\Phi_{CC^{\prime}}$ and cancel the projection onto its orthogonal complement $I_{CC^{\prime}}-\Phi_{CC^{\prime}}$. So after the P-CTC the final state is 
\begin{multline}
\frac{1}{z_0}\left[  p\langle \Phi\vert _{CC^{\prime}}U_{SC}
\left(\rho_{S}\otimes\Phi_{CC^{\prime}}\right)  U_{SC}^{\dag}\left\vert
\Phi\right\rangle _{CC^{\prime}}+\left(  1-p\right)  \left\langle
\Phi\right\vert _{CC^{\prime}}U_{SC}\left(  \rho_{S}\otimes\pi_{C}\otimes
\pi_{C^{\prime}}\right)  U_{SC}^{\dag}\vert \Phi\rangle_{CC^{\prime}}\right]
\otimes\Phi_{CC^{\prime}} \\
=\frac{1}{z_0}\left[  \frac{p}{\left\vert C\right\vert ^{2}}\text{Tr}_{C}
\left\{  U_{SC}\right\}  \rho_{S}\text{Tr}_{C}\left\{  U_{SC}^{\dag}\right\}
+ \frac{\left(  1-p\right)  }{\left\vert C\right\vert ^{3}}
\text{Tr}_{C}\left\{  U_{SC}\left(  \rho_{S}\otimes I_{C}\right)
U_{SC}^{\dag}\right\}  \right]  \otimes\Phi_{CC^{\prime}},
\end{multline}
where $z_0$ is a normalization factor and $|C|$ denotes the dimension of system $C$. By choosing $p=1/\left(  \left\vert C\right\vert +1\right)$, we can ensure that the evolution is equivalent to that given in (\ref{eq:T-CTC-evolve}),  which proves that P-CTCs can simulate T-CTCs.

\subsection{T-CTCs can simulate P-CTCs}

We now show how to simulate a P-CTC with arbitrarily good accuracy using a T-CTC. To improve the accuracy requires a larger CTC system, but the scaling is reasonable.  (We will show that an exponential improvement in accuracy requires only a linearly increasing number of qubits.)

For the simulation, we require $n$ single-qubit systems $C_{1}\cdots C_{n}$ for the CTC systems. To simplify the description, let $\vert \phi^{\rho}\rangle _{RS}$ be a purification of $\rho_{S}$, so that Tr$_{R}\left\{\vert \phi^{\rho}\rangle \langle \phi^{\rho}\vert_{RS}\right\}  =\rho_{S}$. We begin by preparing the state
\begin{equation}
U_{SC}\left[  \vert \phi^{\rho}\rangle _{RS}\vert\Phi\rangle _{CC^{\prime}}\right]  .
\end{equation}
Next, we perform the isometry
\begin{equation}
\vert 0\rangle _{A}\otimes\vert \Phi\rangle \langle
\Phi\vert _{CC^{\prime}}+\vert 1\rangle _{A}\otimes
\left[I_{CC^{\prime}}-\vert \Phi\rangle \langle \Phi\vert_{CC^{\prime}}\right]  ,
\end{equation}
so that the resulting state is
\begin{equation}
\vert 0\rangle _{A}\otimes\vert \Phi\rangle \langle
\Phi\vert _{CC^{\prime}}U_{SC}\left[  \left\vert \phi^{\rho}
\right\rangle _{RS}\vert \Phi\rangle _{CC^{\prime}}\right]
+\vert 1\rangle _{A}\otimes\left[  I_{CC^{\prime}}
-\vert\Phi\rangle \langle \Phi\vert _{CC^{\prime}}\right]
U_{SC}\left[  \vert \phi^{\rho}\rangle _{RS}\left\vert
\Phi\right\rangle _{CC^{\prime}}\right]  .
\label{eq:T-CTC-prep}
\end{equation}
Setting
\begin{align}
p_{0}  &  \equiv\left\Vert \vert \Phi\rangle
\langle\Phi\vert _{CC^{\prime}}U_{SC}\left[  \left\vert
\phi^{\rho}\right\rangle_{RS}\vert \Phi\rangle _{CC^{\prime}}\right]
\right\Vert _{2}^{2},\\
p_{1}  &  \equiv\left\Vert \left[  I_{CC^{\prime}}-\vert \Phi\rangle
\langle \Phi\vert _{CC^{\prime}}\right] U_{SC}
\left[  \vert \phi^{\rho}\rangle _{RS}\left\vert
\Phi\right\rangle_{CC^{\prime}}\right]  \right\Vert _{2}^{2},
\end{align}
we can rewrite the state in (\ref{eq:T-CTC-prep}) as
\begin{equation}
\vert \varphi\rangle _{ARSCC^{\prime}}
\equiv\sqrt{p_{0}}\left\vert0\right\rangle _{A}
\otimes\vert \psi_{0}\rangle _{RSCC^{\prime}}
+\sqrt{p_{1}}\vert 1\rangle _{A}\otimes\vert \psi_{1}\rangle _{RSCC^{\prime}},
\end{equation}
where $\vert \psi_{0}\rangle _{RSCC^{\prime}}$ and
$\left\vert\psi_{1}\right\rangle _{RSCC^{\prime}}$ are normalized states defined by
\begin{align}
\vert \psi_{0}\rangle _{RSCC^{\prime}}  &  \equiv p_{0}^{-1/2}
\vert \Phi\rangle \langle \Phi\vert _{CC^{\prime}}U_{SC}
\left[  \vert \phi^{\rho}\rangle _{RS}\vert\Phi\rangle _{CC^{\prime}}\right]  ,\\
\vert \psi_{1}\rangle _{RSCC^{\prime}}  &  \equiv p_{1}^{-1/2}
\left[  I_{CC^{\prime}}-\vert \Phi\rangle \langle\Phi\vert _{CC^{\prime}}\right] 
U_{SC}\left[  \vert \phi^{\rho}\rangle _{RS}\vert \Phi\rangle _{CC^{\prime}}\right]  .
\end{align}
Observe that the normalization factor $N$ from (\ref{eq:p-ctc-normalization}) is equal to $p_{0}$.

We finally perform the following unitary interaction between the ancilla system $A$ and the $n$ CTC systems $C_{1}\cdots C_{n}$:
\begin{equation}
V_{AC_{1}\cdots C_{n}}\equiv\vert 0\rangle \langle 0\vert _{A}
\otimes I_{C_{1}}\otimes\cdots\otimes I_{C_{n}}
+\vert 1\rangle \langle 1\vert _{A}
\otimes X_{C_{1}}\otimes \cdots \otimes X_{C_{n}}.
\end{equation}
We then use (\ref{eq:T-CTC-evolve}) to calculate the final state:
\begin{multline}
\propto\text{Tr}_{C_{1}\cdots C_{n}}\{  V_{AC_{1}\cdots C_{n}}\}
\vert \varphi\rangle \langle \varphi\vert_{ARSCC^{\prime}}
\text{Tr}_{C_{1}\cdots C_{n}}\{  V_{AC_{1}\cdots C_{n}}^{\dag}\}
\label{eq:T-CTC-simulation}\\
+\text{Tr}_{C_{1}\cdots C_{n}}\{  V_{AC_{1}\cdots C_{n}}(\vert \varphi\rangle
\langle \varphi\vert_{ARSCC^{\prime}}
\otimes I_{C_{1}\cdots C_{n}})  V_{AC_{1}\cdots C_{n}}^{\dag}\}  .
\end{multline}
Since $\text{Tr}_{C_{1}\cdots C_{n}}\{  V_{AC_{1}\cdots C_{n}}\}
=2^{n}\vert 0\rangle \langle 0\vert _{A}$, it follows that\footnote{We note that
\eqref{eq:sat-allen-bound} demonstrates that the vector $\sqrt{p_{0}}\left\vert0\right\rangle _{A}
\otimes\vert \psi_{0}\rangle _{RSCC^{\prime}}$ and the operator 
$\text{Tr}_{C_{1}\cdots C_{n}}\{  V_{AC_{1}\cdots C_{n}}\}$ saturate the bound given in \cite[Eq.~(10)]{A14}. That is, to saturate \cite[Eq.~(10)]{A14}, we can therein set $P = \text{Tr}_{C_{1}\cdots C_{n}}\{  V_{AC_{1}\cdots C_{n}}\}$, $\vert \psi \rangle  = \sqrt{p_{0}}\left\vert0\right\rangle _{A}
\otimes\vert \psi_{0}\rangle _{RSCC^{\prime}}$, the chronology-respecting systems to be $ARSCC^{\prime}$, and the chronology-violating systems to be $C_{1}\cdots C_{n}$. We thank John-Mark Allen for pointing this out to us.}
\begin{equation}
\text{Tr}_{C_{1}\cdots C_{n}}\{  V_{AC_{1}\cdots C_{n}}\}
\vert \varphi\rangle \langle \varphi\vert_{ARSCC^{\prime}}
\text{Tr}_{C_{1}\cdots C_{n}}\{  V_{AC_{1}\cdots C_{n}}^{\dag}\}
=2^{2n}p_{0}\vert 0\rangle \langle 0\vert _{A}
\otimes\vert \psi_{0}\rangle \langle \psi_{0}\vert _{RSCC^{\prime}},\label{eq:sat-allen-bound}
\end{equation}
\begin{multline}
\text{Tr}_{C_{1}\cdots C_{n}}\{  V_{AC_{1}\cdots C_{n}}
(  \vert\varphi\rangle \langle \varphi\vert _{ARSCC^{\prime}}
\otimes I_{C_{1}\cdots C_{n}})  V_{AC_{1}\cdots C_{n}}^{\dag}\}
=\\2^{n}p_{0}\vert 0\rangle \langle 0\vert_{A}
\otimes\vert \psi_{0}\rangle \langle \psi_{0}\vert_{RSCC^{\prime}}
+2^{n}p_{1}\vert 1\rangle \langle 1\vert_{A}
\otimes\vert \psi_{1}\rangle \langle \psi_{1}\vert_{RSCC^{\prime}}.
\end{multline}
Thus, after normalization, (\ref{eq:T-CTC-simulation}) becomes%
\begin{equation}
\frac{\left(  2^{n}+1\right)  p_{0}}{\left(  2^{n}+1\right)  p_{0}+p_{1}%
}\vert 0\rangle \langle 0\vert _{A}\otimes\vert
\psi_{0}\rangle \langle \psi_{0}\vert _{RSCC^{\prime}}%
+\frac{p_{1}}{\left(  2^{n}+1\right)  p_{0}+p_{1}}\vert 1\rangle
\langle 1\vert _{A}\otimes\vert \psi_{1}\rangle
\langle \psi_{1}\vert _{RSCC^{\prime}}.
\end{equation}
With only a linear increase in the number of CTC qubits $n$, we get an exponential suppression of the undesired second term, and achieve an arbitrarily good approximation to the desired P-CTC evolution given in (\ref{eq:P-CTC-evolve}).

\section{Simulating D-CTCs}
\label{sec:D-CTC-simulation}

We now discuss a few methods for quantum computers to simulate D-CTCs. The approach outlined here has a guaranteed fixed-point convergence and thus can be viewed as an alternative to the equivalent circuit model \cite{RM10,RD12}. We should clarify at the outset that the simulation is such that any third party would not be able to distinguish the output of a D-CTC-assisted circuit from our simulation. However, the proposed simulation need not be efficient:  it could take a long time due to the need for convergence to a fixed point of a quantum circuit, the computation of which is known to be a PSPACE-complete problem \cite{AW09}. Note that this simulation method does not involve classically solving the Deutsch consistency condition as in \cite{RBMWR14}.  Thus, the simulation we propose here would be more interesting in our opinion, though far more experimentally challenging.

Recall from Section~\ref{sec:D-CTC-review}\ that the initial state of the chronology-respecting system is $\rho_{S}$, and the unitary interaction is $U_{SC}$. To simulate a D-CTC, it is clear that one could simulate the fixed-point projection of (\ref{eq:fixed-point-approx}) where $N$ is a large positive integer. To this end, we prepare a control register $F_{1}\cdots F_{N}$ of $N$ qubits in a uniform superposition of Dicke states \cite{D54} of the following form:
\begin{equation}
\frac{1}{\sqrt{N+1}}\sum_{k=0}^{N}\left\vert D_{k}^{N}\right\rangle ,
\end{equation}
where
\begin{equation}
\left\vert D_{k}^{N}\right\rangle \equiv\frac{1}{\sqrt{\binom{N}{k}}}
\sum_{j}C_{j}\{\vert 0\rangle ^{\otimes k}\otimes\vert 1\rangle ^{\otimes N-k}\},
\end{equation}
and the second sum is over all combinations of
$\vert 0\rangle^{\otimes k}\otimes\vert 1\rangle ^{\otimes N-k}$
(denoted by $C_{j}\{\vert 0\rangle ^{\otimes k}\otimes\vert 1\rangle^{\otimes N-k}\}$).  Next, we initialize $N$ registers $S_{1}\cdots S_{N}$ to $N$ copies of the state $\rho_{S}$.  This requirement that we are able to prepare multiple copies of $\rho_{S}$ is necessary to simulate a D-CTC with an ordinary quantum computer---otherwise simulation is impossible.  Let $C_{1}$ denote the first CTC system---initialized in some state $\omega_{C_{1}}$. The simulation
proceeds by applying the following controlled unitary to registers $F_{1}$, $S_{1}$, and $C_{1}$:
\begin{equation}
\vert 0\rangle \langle 0\vert _{F_{1}}\otimes
I_{S_{1}C_{1}}+\vert 1\rangle \langle 1\vert _{F_{1}}\otimes U_{S_{1}C_{1}}.
\end{equation}
Let $C_{2}$ denote the output CTC system (in general, we let $C_{i+1}$ denote the CTC output of the $i$th controlled unitary). The rest of the simulation proceeds as follows:

\begin{enumerate}
\item Set $i=2$.

\item Apply the following controlled unitary to systems $F_{i}$, $S_{i}$, and $C_{i}$:
\begin{equation}
\vert 0\rangle \langle 0\vert _{F_{i}}\otimes
I_{S_{i}C_{i}}+\vert 1\rangle \langle 1\vert _{F_{i}}
\otimes U_{S_{i}C_{i}}.
\label{eq:ith-controlled-U}
\end{equation}

\item Set $i:=i+1$. If $i\leq N$, go to step 2. Otherwise, output system $S_{N}$ (while discarding all other systems).
\end{enumerate}

\begin{figure}[ptb]
\begin{center}
\includegraphics[
width=3.531in
]{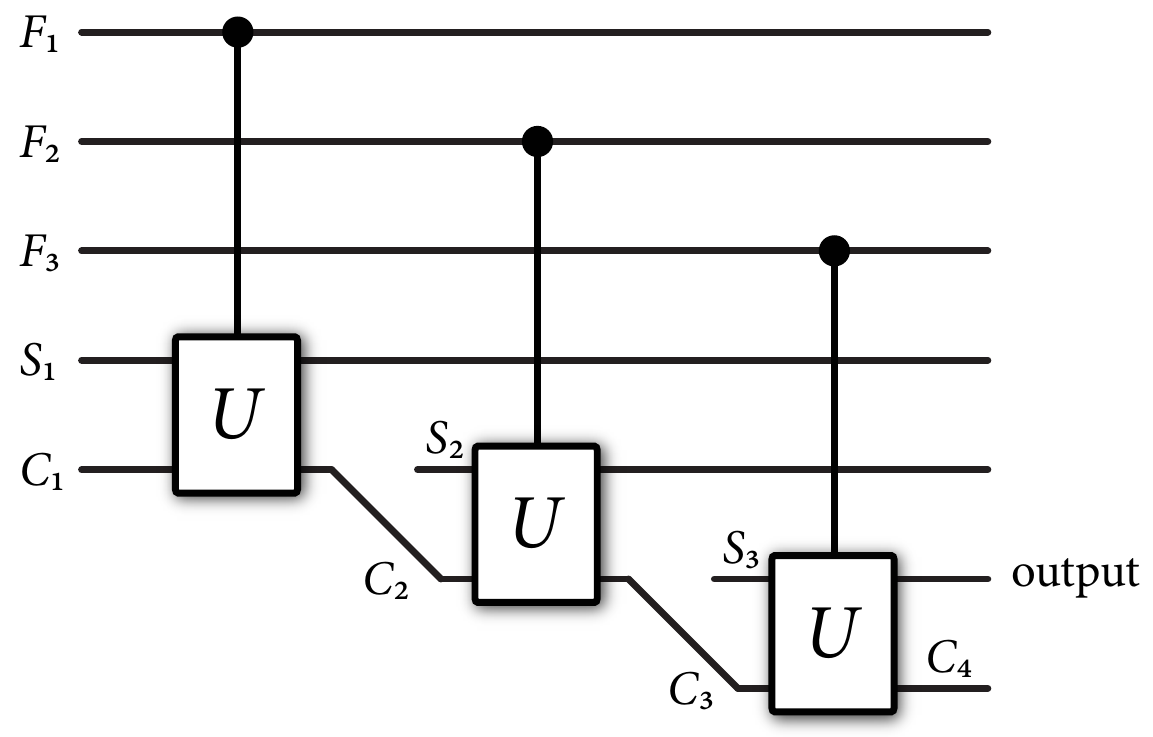}
\end{center}
\caption{This figure depicts a D-CTC simulation for which $N=3$.}
\label{fig:D-CTC-simulation}%
\end{figure}

Figure~\ref{fig:D-CTC-simulation} depicts the simulation. $N$ needs to be large enough to ensure convergence. In practice, for many choices of $\rho_{S}$ and $U_{SC}$, $N$ will be prohibitively large.

The result of the procedure is the following evolution:
\begin{equation}
\text{Tr}_{C}\left\{  U_{SC}\left(  \rho_{S}
\otimes\overline{\omega_{C}^{N}}\right)  U_{SC}^{\dag}\right\}  ,
\end{equation}
where
\begin{equation}
\overline{\omega_{C}^{N}}\equiv\overline{\mathcal{N}_{U,\rho}^{N}}
\left(\omega_{C}\right)  .
\end{equation}
The idea is that one could choose $N$ large enough such that
\begin{equation}
\left\Vert \text{Tr}_{C}\left\{  U_{SC}\left(  \rho_{S}
\otimes\overline{\omega_{C}^{N}}\right)  U_{SC}^{\dag}\right\}
- \text{Tr}_{C}\left\{U_{SC}\left(  \rho_{S}\otimes\sigma_{C}\right)  U_{SC}^{\dag}\right\}
\right\Vert _{1}\leq\varepsilon
\end{equation}
for a given accuracy $\varepsilon>0$, where
$\sigma_{C}=\overline{\mathcal{N}_{U,\rho}^{\infty}}(  \omega_{C})$.

One could employ other states of the control register, such as the following
state:
\begin{equation}
\frac{1}{\sqrt{N}}\left(  \left\vert 000\cdots0\right\rangle
+ \left\vert100\cdots0\right\rangle +\left\vert 110\cdots0\right\rangle
+ \left\vert111\cdots0\right\rangle +\cdots+\left\vert 111\cdots1\right\rangle \right)  ,
\end{equation}
and the outcome would be the same as the above.

We could also employ a different ``iterate'' for the fixed-point approximation, which might be easier to implement experimentally while having a faster rate of convergence. This other approach (called Krasnoselskij iteration \cite{B07}) amounts to initializing the control register $F_{1}\cdots F_{N}$ as follows:%
\begin{equation}
\bigotimes\limits_{i=1}^{N}\left[  \sqrt{\gamma}\vert 0\rangle_{F_{i}}
+ \sqrt{1-\gamma}\vert 1\rangle _{F_{i}}\right]  ,
\end{equation}
where $\gamma\in\left(  0,1\right)$ is a constant. The rest of the simulation proceeds as in the steps given above, leading to the following fixed-point approximation for the state of the CTC system:
\begin{equation}
\mathcal{N}_{U,\rho}^{\gamma,N}(  \omega_{C})
\equiv \underbrace{\left(  \mathcal{N}_{U,\rho}^{\gamma}\circ\cdots\circ
\mathcal{N}_{U,\rho}^{\gamma}\right)  }_{N\text{ times}}(  \omega_{C})  ,
\end{equation}
where%
\begin{equation}
\mathcal{N}_{U,\rho}^{\gamma}(  \omega_{C})  \equiv\gamma\omega
_{C}+\left(  1-\gamma\right)  \mathcal{N}_{U,\rho}(  \omega_{C})  .
\end{equation}
By taking the limit as $N\rightarrow\infty$, we find that for all $\gamma\in\left(  0,1\right)  $ (see, e.g., \cite[Theorem~3.2]{B07})
\begin{equation}
\lim_{N\rightarrow\infty}\mathcal{N}_{U,\rho}^{\gamma,N}(  \omega_{C})
= \overline{\mathcal{N}_{U,\rho}^{\infty}}(  \omega_{C})  .
\end{equation}
That is, the limiting map converges to the projection onto the fixed points of $\mathcal{N}_{U,\rho}$. Thus, this procedure works equally well in simulating D-CTCs and has the advantage that the control register might be easier to prepare, as it is a tensor-product state.

\subsection{Maximum entropy CTC state}

The last aspect that we address is Deutsch's maximum entropy postulate.  That is:  what initial state of the register $C_{1}$ in the above simulation will guarantee that the limiting state of the CTC is a maximum entropy state?
An initial guess is that the initial state should be maximally mixed, but this turns out to be correct only in some cases.

We now show how to choose the initial state to make the state of the CTC a maximum entropy state. We make use of (\ref{eq:fixed-point-projection}) and the formula for quantum entropy $H(  \rho)  \equiv-\operatorname{Tr}\{  \rho\log\rho\}$ for a density operator $\rho$ with the logarithm taken to have base two. When we input a state $\omega_{C}$ to $\overline{\mathcal{N}_{U,\rho}^{\infty}}$, the output is
\begin{equation}
\bigoplus\limits_{j}\text{Tr}_{C_{R_{j}}}\{\Pi_{j}\omega_{C}\Pi_{j}%
\}\otimes\rho_{j}.\label{eq:fixed-point-state-MAX-ENT}%
\end{equation}
Let
\begin{equation}
\omega_{j}  \equiv\frac{1}{q(j)}\text{Tr}_{C_{R_{j}}}\{\Pi_{j}\omega_{C}\Pi_{j}\},\ \ \ 
q(j)  \equiv\text{Tr}\{\Pi_{j}\omega_{C}\Pi_{j}\},
\end{equation}
so that we can rewrite (\ref{eq:fixed-point-state-MAX-ENT}) as
\begin{equation}
\bigoplus\limits_{j}q(j)\ \omega_{j}\otimes\rho_{j}.
\end{equation}
The quantum entropy of the above state is
\begin{equation}
H(  \{q(j)\})  +\sum_{j}q(j)\left[  H(  \omega_{j})
+H(  \rho_{j})  \right]  ,
\end{equation}
where $H(  \{q(j)\})  \equiv-\sum_{j}q(j)\log q(j)$ is the Shannon entropy of the distribution $q$. It is clear that we should pick $\omega_{j}=\pi_{j}$ (the maximally mixed state) because $H(  \omega_{j}) \leq H(  \pi_{j})  =\log d_{L_{j}}$. So our goal is to optimize the following objective function with respect to the distribution $q(j)$:
\begin{equation}
H(  \{q(j)\})  +\sum_{j}q(j)\left[  \log d_{L_{j}} + H(\rho_{j})  \right]  .
\end{equation}
We employ the method of Lagrange multipliers. Our Lagrangian is
\begin{equation}
H(  \{q(j)\})  +\sum_{j}q(j)\left[  \log d_{L_{j}}
+ H(\rho_{j})  \right]  +\lambda\left[  \sum_{j}q(j)-1\right]  ,
\end{equation}
where $\lambda\in\mathbb{R}$ is a Lagrange multiplier. The derivative with respect to $q(j)$ is
\begin{equation}
-\log q(j)-1+\log d_{L_{j}}+H(  \rho_{j})  +\lambda.
\end{equation}
Setting this equal to zero, and solving for $q(j)$ gives
\begin{equation}
q(j)=2^{\lambda}2^{H(  \rho_{j})  +\log d_{L_{j}}-1}.
\end{equation}
We then choose $\lambda$ to normalize the $q(j)$'s such that $q$ is a probability distribution:
\begin{equation}
\lambda=-\log\left(  \sum_{j}2^{H(  \rho_{j})  +\log d_{L_{j}}-1}\right)  .
\end{equation}
Thus, we can take the initial state of the CTC system to be
\begin{equation}
\sigma_{C}\equiv\bigoplus\limits_{j}q(j)\ \pi_{j}\otimes\rho_{j},
\end{equation}
with $q(j)$ as given above, to guarantee that the state of the CTC is a maximum-entropy state. Also, one can easily check for this choice that
\begin{equation}
\bigoplus\limits_{j}\text{Tr}_{C_{R_{j}}}\{\Pi_{j}\sigma_{C}\Pi_{j}%
\}\otimes\rho_{j}=\sigma_{C}.
\end{equation}
Presumably, for many choices of chronology-respecting states $\rho_{S}$ and interaction unitaries $U_{SC}$, it should be computationally difficult to prepare the state of the CTC system as given above.  Otherwise, one would be able to solve problems in PSPACE efficiently. However, one can observe that if the second factor in each block of the direct sum is trivial, so that $\rho_j = 1$, and if the size of each block is the same, then the maximum-entropy state of the CTC system is the maximally mixed state. For these special cases, the preparation of the CTC system would of course be computationally efficient, but the resulting D-CTC circuit then cannot accomplish any information-processing task that is not possible with an ordinary quantum channel (since, in that case, an ordinary quantum channel can easily simulate this D-CTC). 

\section{Conclusion}

This paper addresses the simulation of CTC-enabled evolutions in two different ways. First, we prove that the recently introduced T-CTCs are physically equivalent to P-CTCs. Next, we showed how quantum computers with many copies of the initial input state can simulate D-CTCs, though not necessarily efficiently. We also addressed some ambiguities associated with the equivalent circuit model \cite{RM10,RD12}, and we determined an explicit form for the state that converges to the maximum entropy state of a D-CTC.

There is an interesting and potentially practical consequence of our work here. Is it possible to employ the method outlined in Section~\ref{sec:D-CTC-simulation} and the state discrimination circuit of \cite{BHW09} to produce a useful method for distinguishing non-orthogonal states if one has $N$ copies available? To determine if the method would be competitive with existing algorithms that use a constant-sized quantum memory \cite{BCZ13}, one would need to determine convergence rates of the method outlined in Section~\ref{sec:D-CTC-simulation}. We leave this as an open question for the interested reader.

\bigskip

\textbf{Acknowledgements.} We are especially grateful to Tom Cooney for many enlightening discussions on fixed points of CPTP linear\ maps. We thank John-Mark Allen for helpful feedback that improved the manuscript. We also acknowledge Jonathan Dowling for his help in obtaining FQXI\ funds to support this research. Finally, we acknowledge support from the Department of Physics and Astronomy at Louisiana State University and the Foundational Questions Institute (FQXI) for supporting the grant ``Closed timelike curves and quantum information processing.''

\bibliographystyle{alpha}
\bibliography{Ref}

\end{document}